\documentclass[runningheads]{llncs}
\usepackage[T1]{fontenc}
\usepackage{graphicx}
\usepackage{booktabs}
\usepackage[misc]{ifsym}
\newcommand{\corr}{(\Letter)}
\usepackage{mwe}
\usepackage{hyperref}
\usepackage{enumitem}
\usepackage{amsmath}
\usepackage{amssymb}
\usepackage{multirow}
\usepackage{xcolor}

\begin{document}

\title{ReFORM: Review-aggregated Profile Generation via LLM
with Multi-Factor Attention for Restaurant Recommendation}

\titlerunning{ReFORM}

\author{Moonsoo Park\inst{1} \and
Seulbeen Je\inst{2} \and
Donghyeon Park\inst{2} \corr}

\authorrunning{M. Park et al.}

\institute{University of Southern California, USA, \email{moonsoo@usc.edu}
\and
Sejong University, South Korea, \email{powerjsv@sju.ac.kr, parkdh@sejong.ac.kr}}

\maketitle              

\begin{abstract}
In recommender systems, large language models (LLMs) have gained popularity for generating descriptive summarization to improve recommendation robustness, along with Graph Convolution Networks. However, existing LLM-enhanced recommendation studies mainly rely on the internal knowledge of LLMs about item titles while neglecting the importance of various factors influencing users' decisions. Although information reflecting various decision factors of each user is abundant in reviews, few studies have actively exploited such insights for recommendation. To address these limitations, we propose a \textbf{ReFORM}: \textbf{Re}view-aggregated Profile Generation via LLM with Multi-\textbf{F}act\textbf{O}r Attentive \textbf{R}eco\textbf{M}mendation framework. Specifically, we first generate factor-specific user and item profiles from reviews using LLM to capture a user's preference by items and an item's evaluation by users. Then, we propose a Multi-Factor Attention to highlight the most influential factors in each user's decision-making process. In this paper, we conduct experiments on two restaurant datasets of varying scales, demonstrating its robustness and superior performance over state-of-the-art baselines. Furthermore, in-depth analyses validate the effectiveness of the proposed modules and provide insights into the sources of personalization. Our source code and datasets are available at \href{https://github.com/m0onsoo/ReFORM}{\url{https://github.com/m0onsoo/ReFORM}}.

\end{abstract}

\section{Introduction}
Recommender systems play a vital role in filtering vast amounts of information, offering personalized content recommendations across domains such as e-commerce \cite{onlineshopping}, movies \cite{movie_cf}, and social networks \cite{graph_social}. Collaborative filtering (CF), which models user-item interactions to predict future preferences, serves as the foundation of many recommender systems.
Notably, Graph Convolution Networks (GCNs) have emerged as a powerful paradigm in collaborative filtering, enabling the aggregation of high-order collaborative signals by representing users and items as nodes in a bipartite graph \cite{ngcf,gccf,lgcn,sgl,simgcl}. Nevertheless, these methods still rely exclusively on interaction data, overlooking valuable external feedback (e.g., ratings and user reviews) that offers richer decision reasoning.

\begin{figure}[!t]
  \centering
  \includegraphics[width=2.5in]{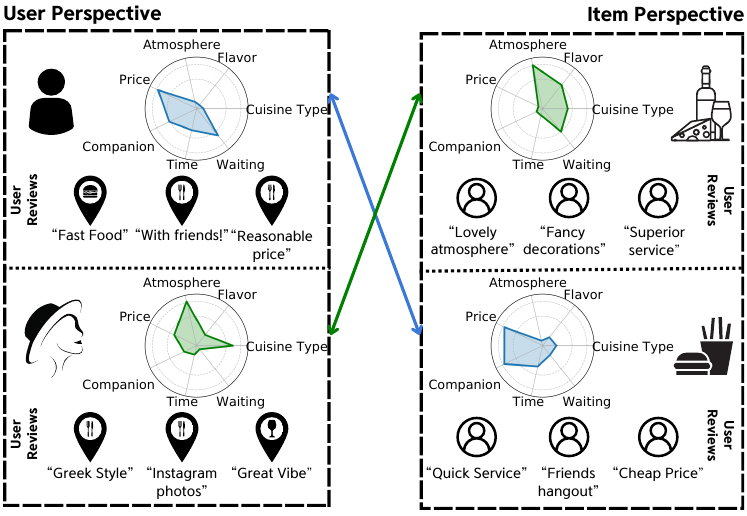}
  \caption{An example of capturing different users’ personal preferences and restaurants’ prominent characteristics based on their reviews. By matching a user's preferences and a restaurant's characteristics, our ReFORM framework provides more personalized and robust recommendations.}
  \label{figure:toy}
\end{figure}

Mining hidden factors from review text helps overcome the limitations of interaction-based recommender systems \cite{rst_context,factor_movie}, as reviews offer detailed context on user preferences and item attributes that are absent in binary interactions. By leveraging these insights, models can build more expressive and personalized representations \cite{rlmrec}. Recent efforts increasingly exploit LLMs \cite{gpt4,llama} for advanced text understanding and generation, extracting deeper insights from interaction history and user reviews. Specifically, existing LLM-enhanced recommendation methods (e.g., KAR \cite{kar}, RLMRec \cite{rlmrec}) primarily focus on aligning general semantic features with graph structures or open-world knowledge. However, they still lack the granular, personalized modeling of diverse decision factors that are vividly expressed in individual user reviews. 

To overcome the aforementioned challenges, we propose a \textbf{ReFORM}: \textbf{Re}view\allowbreak-aggregated Profile Generation via LLM with Multi-\textbf{F}act\textbf{O}r Attentive \textbf{R}eco\allowbreak\textbf{M}mendation framework that 1) generates factor-specific user and item profiles exclusively from reviews through Review-aggregated Profile Generation (RPG) and 2) captures nuanced factor preferences from the profiles through Multi-Factor Attention (MFA) mechanism.




As illustrated in Figure \ref{figure:toy}, ReFORM aims to identify the specific factor preferences driving user decisions. It consists of two main components. First, Review-aggregated Profile Generation (RPG) utilizes LLMs to distill unstructured reviews into explicit, factor-specific profiles for both users and items. Second, rather than treating all factors equally, the Multi-Factor Attention (MFA) mechanism dynamically highlights the most influential factors in each user's decision-making process by computing attention across user-item interactions. Finally, these factor-attentive embeddings are integrated with standard graph node embeddings for precise and personalized recommendations.

Extensive experiments validate that our two main approaches, Review-aggr\-egated Profile Generation (RPG) and Multi-Factor Attention (MFA), outperform state-of-the-art baselines in both performance and robustness. In addition, we conduct in-depth analyses, including factor-wise ablation to quantify the contribution of each factor and noise injection experiments in the RPG stage to assess the importance of authentic user reviews and LLM-based profiling. The major contributions are summarized as follows:
\begin{itemize}[leftmargin=1em]
    \item We propose ReFORM, a novel LLM-enhanced recommendation framework. It uniquely combines Review-aggregated Profile Generation to extract factor-specific profiles from reviews, and a Multi-Factor Attention mechanism to dynamically capture the most influential decision factors for each user. 
    \item ReFORM framework demonstrates extensive experiments that our method significantly outperforms existing baselines, validating its effectiveness in improving recommendation performance.
    \item We empirically validate the core design of ReFORM through factor ablation and review-noise injection studies, demonstrating that the selected factors and review-derived profiles capture essential personalized signals for effective recommendation.
\end{itemize}
\section{Related Work}

\subsection{Graph-based Collaborative Filtering}

Graph-based collaborative filtering (CF) has been widely adopted to model relationships between users and items through interactions, leveraging graph convolution networks to improve recommendation accuracy. Building on this line of research, methods such as NGCF \cite{ngcf}, GCCF \cite{gccf}, and LightGCN \cite{lgcn} have pushed the boundaries of graph-based collaborative filtering. However, these methods still face challenges due to the sparsity and noise in implicit feedback data. To address these challenges, self-supervised learning has been introduced to mitigate the limitations of GCNs and enhance graph-based CF models \cite{sgl,simgcl}. However, graph-based recommendation systems primarily focus on aggregating structural relationships, making it difficult to leverage text-based insights that explain the reasons behind user-item interactions. While models like PinSage \cite{pinsage}, MMGCN \cite{mmgcn}, and GRCN \cite{grcn} incorporate multi-modal features for user preference modeling, they are limited by their reliance on item-level information. In real-world scenarios, however, users make decisions based on diverse contextual factors, highlighting the need to capture user-level information as well.

Prior studies \cite{rst_context,factor_movie} have shown that contextual information from reviews improves recommendation quality. Motivated by these findings, we extract decisive textual factors and integrate them into GCNs at both the user and item levels to enhance personalization.

\begin{figure*}[!htbp]
  \centering
  \includegraphics[width = 4.6in]{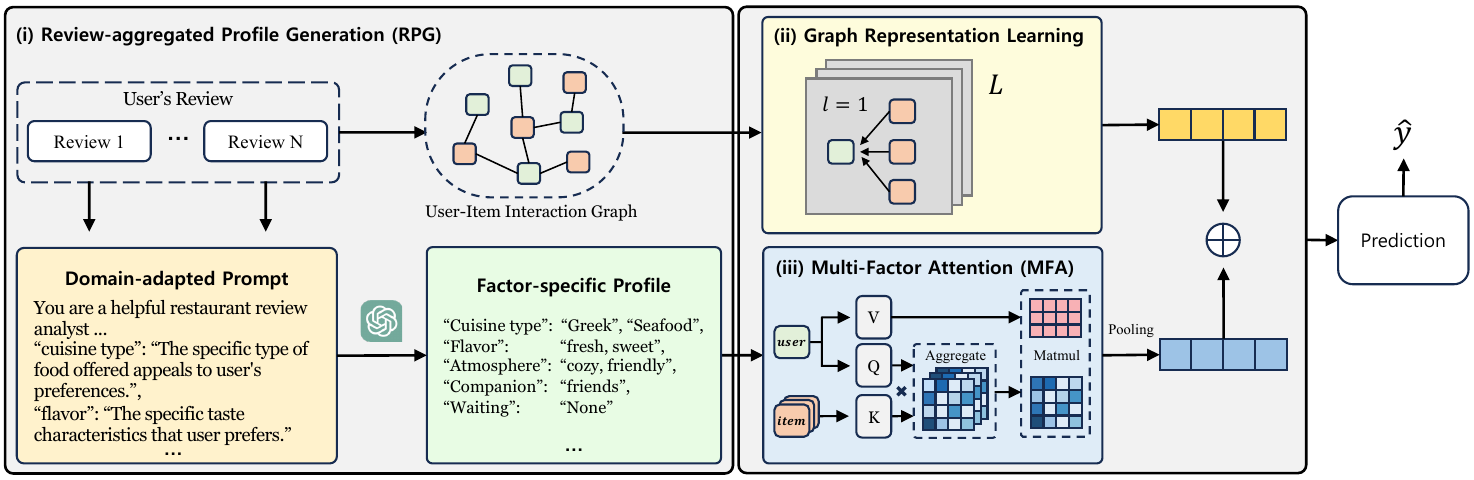}
  \caption{The overall framework of ReFORM. (i) Review-aggregated Profile Generation (RPG) constructs factor-specific user and item profiles from reviews based on domain-adapted prompt. (ii) Graph Representation Learning captures high-order interactions. (iii) Multi-Factor Attention highlights the most influential factors in each user's decision-making process for better recommendation results.}
  \label{figure:framework}
\end{figure*}

\subsection{Large Language Models for Recommendation}

Large language models (LLMs) have gained significant interest in Recommendation systems (RS) due to their advanced text understanding and generation capabilities. Several studies have explored the use of LLMs as inference models by fine-tuning language models for recommendation tasks \cite{llm_inference1,llm_inference2,tallrec}. For instance, P5 \cite{p5} reformulates user interaction data into natural language sequences, enabling the training of language models for recommendations. However, directly employing LLMs as a recommender shows limited performance compared to traditional RSs \cite{llmasrec_limit1,llmasrec_limit2,llmasrec_limit3}. 

Beyond using LLMs as stand-alone recommendation systems, some studies \cite{llm-rec,llmrec,llm4rec_desc1,llm4rec_desc2} have employed LLMs to generate enhanced text features for traditional recommender systems. Notably, KAR \cite{kar} used scenario-specific factors to deduce user preferences from generic metadata (e.g., movie titles or genres), effectively integrating open-world knowledge into recommender systems. Despite the advancement, relying heavily on internal knowledge of LLMs may fail to capture the nuanced aspects of personal user experiences (e.g., atmosphere, companion) found in user reviews. Also, RLMRec \cite{rlmrec} shows its robustness in integrating semantic representations from metadata and reviews with various GCN methods. However, treating all semantic information equally may introduce noise and degrade recommendation quality \cite{mlp_noise} for certain users.

In this paper, to mitigate these challenges, we extract domain-adapted factors solely from reviews through Review-aggregated Profile Generation (RPG) using LLM. These factor-specific profiles are further modeled through Multi-Factor Attention (MFA) to finely adjust the importance of factors for each individual.

\section{Methodology}

\subsection{Overview}
The proposed framework is illustrated in Figure \ref{figure:framework}. It begins with Review-aggre\allowbreak-gated Profile Generation (RPG) by generating user and item profiles enriched with aggregated reviews and domain-adapted factor prompts, which utilizes a large language model (LLM). The model then learns from both graph-based representation and profile-based representation to enhance recommendation performance. Our Multi-Factor Attention (MFA) mechanism utilizes pre-generated profiles of RPG to capture the factor importance of both user-to-items and item-to-users perspectives. Finally, the two representations are integrated to produce Top-\(K\) recommendation results.

\subsection{Graph Representation Learning}
The proposed framework leverages the propagation rule of LightGCN \cite{lgcn}, extracting node-level representations through graph convolution without feature transformation and nonlinear activation. During the propagation for the \( l \) layers, the model aggregates information from neighboring nodes to refine graph embeddings \( \mathbf{e}^g \). This process captures high-order collaborative signals while maintaining computational simplicity, as illustrated in Figure \ref{figure:framework}-(ii). Here, \(\mathcal{N}_u\) and \( \mathcal{N}_i \) denote neighbor node sets of user \( u \) and item \( i \).

\begin{gather}\label{eq:lgcn}
    \mathbf{e}_u^{(l+1)} = \sum_{i \in \mathcal{N}_u} \frac{1}{\sqrt{|\mathcal{N}_u|} \sqrt{|\mathcal{N}_i|}} \mathbf{e}_i^{(l)} ; \quad \mathbf{e}_i^{(l+1)} = \sum_{u \in \mathcal{N}_i} \frac{1}{\sqrt{|\mathcal{N}_i|} \sqrt{|\mathcal{N}_u|}} \mathbf{e}_u^{(l)} \\
    \mathbf{e}_u^g = \sum_{l=1}^{L} \mathbf{e}_u^{(l)} ; \quad 
    \mathbf{e}_i^g = \sum_{l=1}^{L} \mathbf{e}_i^{(l)}
\end{gather}

\subsection{Review-aggregated Profile Generation}

\subsubsection{Restaurant-scenario factors.} People choose restaurants based on a variety of criteria. To define key factors involved in restaurant recommendation, we systematically selected restaurant-scenario factors based on prior studies \cite{rst_context,kar}. Specifically, we consider \textit{cuisine type, flavor, atmosphere, price, time, waiting, and companion}, which reflect the characteristics of the restaurant domain. 

\subsubsection{Factor-specific Profile Generation.} In Review-aggregated Profile Generation (RPG) step, we reform the reviews into detailed factor-specific user/item profiles using a large language model (LLM). We first construct a domain-adapted prompt (Figure \ref{figure:framework}-(i)) that illustrates \( m \) domain-specific factors (\textit{cuisine type, flavor, atmosphere}, etc.). We then query the GPT-4o mini \cite{openai2024gpt4o} model with the aggregated reviews to generate factor-specific profiles.

We sample \( N = 100 \) reviews as input for LLM. For user-level profile construction, we randomly sample reviews, assuming that individual users tend to express consistent preferences throughout their reviews. In contrast, for item-level profile construction, longer reviews are prioritized during sampling. This strategy is motivated by the observation that longer texts tend to exhibit higher lexical coverage and lower distributional variance \cite{textcoverage}, enabling the LLM to capture more comprehensive and nuanced item characteristics. For all the sampled reviews \( r_j \), we provide the domain-adapted prompt to generate descriptions for each factor \( m \). The final user/item profile \( P_{u/i} \) is constructed by aggregating the descriptions across all \( M \) factors:
\begin{equation}\label{eq:prompt}
    P_{u/i} = \{f_m\}_{m=1}^M, \quad f_m = \operatorname{LLM}(\operatorname{Prompt}_m, \{r_j\}_{j=1}^N)
\end{equation}
where \( f_m \) represents the description generated for factor \( m \).

Figure \ref{figure:framework}-(i) illustrates an example of the prompt and the actual response generated by the LLM. In order to generate a profile of a specific user, a sampled subset of reviews of the user and a prompt containing descriptions of factors are provided to guide the LLM in understanding the user's factor-specific preferences. For instance, in the case of \textit{cuisine type}, the instruction \textit{"The specific types of food offered appeal to users' preferences and dietary restrictions."} is included in the prompt, prompting the LLM to extract relevant preferences such as \textit{Greek} and \textit{Seafood} from the user's reviews. By enabling factor-level preference inference directly from user-written reviews, RPG allows for the construction of more personalized and fine-grained user profiles, ensuring a robust and efficient means of
capturing personal preferences for further recommendation tasks.

\subsection{Multi-Factor Attention}
\begin{figure}[!t]
  \centering
  \includegraphics[width = 2.5in]{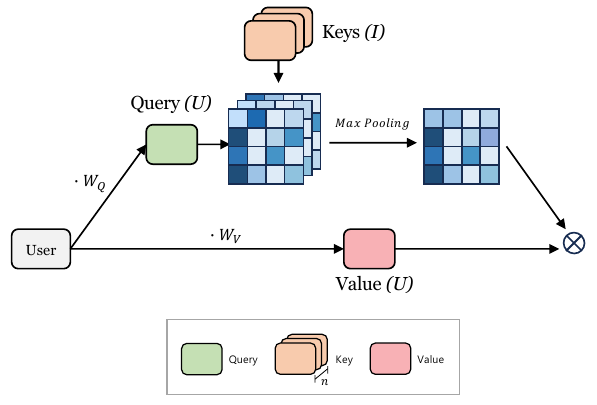}
  \caption{Multi-Factor Attention (MFA) Architecture.}
  \label{figure:MFA}
\end{figure}

The profile information generated by the LLM demonstrates high quality; however, assuming equal importance across all factors hinders the model’s ability to capture user preferences in detail. For more effective recommendations, it is essential to identify which factors a user prioritizes during the decision-making process to ensure better personalization. To this end, we propose a Multi-Factor Attention (MFA) to model factor-level importance, as illustrated in Figure \ref{figure:MFA}.
\subsubsection{Profile Encoding.} We first encode user and item profiles, generated from our previous RPG process, using BERT \cite{bert}. Each factor is transformed into a textual embedding separately to form the profile matrices $\textbf{X}_{P_u}, \textbf{Y}_{P_i} \in \mathbb{R}^{M \times d}$. Here, \( M \) and \( d \) denote the number of factors and the dimension size of the embedding.

\subsubsection{Multi-level Cross Attention.} Typical Attention mechanisms assume \textbf{K} = \textbf{V}, where the encoder's hidden states serve as keys \textbf{K} and values \textbf{V}, and the decoder's hidden states act as queries \textbf{Q} \cite{mt_qkv}. 
In our MFA process, we assume \textbf{Q} = \textbf{V}. That is, user representations serve as the query \textbf{Q}, while item representations serve as the key \textbf{K}. The value \textbf{V}, again derived from user representations, helps refine the nuanced preferences for each specific factor.
\begin{gather}
    \textbf{Q}_u = \textbf{X}_{P_u}\textbf{W}_{Q_u} \quad \textbf{K}_i = \textbf{Y}_{P_i}\textbf{W}_{K_i} \quad \textbf{V}_u = \textbf{X}_{P_u}\textbf{W}_{V_u} \quad \\
    \textbf{Q}_i = \textbf{X}_{P_i}\textbf{W}_{Q_i} \quad \textbf{K}_u = \textbf{Y}_{P_u}\textbf{W}_{K_u} \quad \textbf{V}_i = \textbf{X}_{P_i}\textbf{W}_{V_i} \quad 
\end{gather}
For simplicity, we describe the process from the user's perspective in this section and Figure \ref{figure:MFA}; however, the reverse—items as \textbf{Q} and \textbf{V}, and users as \textbf{K}—is also modeled. The user profile matrix $\textbf{X}_{P_u}$ is transformed into the  query and value representations via the learned projection matrices $\textbf{W}_{Q_u}$ and $\textbf{W}_{V_u}$, respectively, while the item profile matrix $\textbf{Y}_{P_i}$ is transformed into key representations via $\textbf{W}_{K_i}$. The MFA equation is denoted as:
\begin{equation}
    \operatorname{\textbf{MFA}}(\textbf{Q}, \textbf{K}^{1:n}, \textbf{V}) = \max_{i=1}^n (
    \operatorname{softmax}\left(\frac{\textbf{Q} \textbf{K}_i^T}{\sqrt{d^*}}\right) ) \textbf{V} 
    \label{eq:MFA_formula}
\end{equation}
Here, we apply cross attention \cite{transformer} to $\ \textbf{Q}, \textbf{K}^{1:n}, \textbf{V}$  (all of dimension d*) to model the importance of each factor influencing a user's choices on certain items. Specifically, if users tend to choose restaurants based on \textit{cuisine type}, MFA assigns greater weight to that factor. At each epoch, we randomly select \( n \) items from their behavior history and calculate attention weights between user query $\textbf{Q}_u$ and item keys $\textbf{K}_i^{1:n}$ \cite{MVA}. Calculating attention with a single key may bias the model toward a user's sudden decisions, potentially misrepresenting their genuine preferences. To improve robustness, we compare multiple keys to achieve consistent and less biased attention outputs.

\subsubsection{Attention Map Pooling.}  Inspired by the concept of squeeze-and-excitation \cite{GENet,SENet}, which enhances feature representation by capturing critical contextual information through the compression of each feature map, we adopt a simplified approach. Our MFA method aggregates attention maps across multiple keys by applying max pooling to activate specific factors of the values. Max pooling method is to maximize the focus on dominant factors that users are interested in. The aggregated attention map between the query and multiple keys is trained to strongly reflect the user's preferred factors, ensuring that relevant factors are emphasized and effectively integrated into the value representation. The resulting factor-weighted values are a matrix of size  \( M \times d^*\), reflecting the number of factors \( M \) and the embedding dimensions \( d^* \). To obtain the final Multi-Factor Attentive embedding that captures factor-specific preferences, we averaged these weighted value matrices:
\begin{gather}
    \mathbf{e}_u^a = \frac{1}{M}\sum_{m=1}^M \operatorname{\textbf{MFA}}(\textbf{Q}_u ,\textbf{K}_i^{1:n}, \textbf{V}_u ) ; \quad
    \mathbf{e}_i^a = \frac{1}{M}\sum_{m=1}^M \operatorname{\textbf{MFA}}(\textbf{Q}_i ,\textbf{K}_u^{1:n}, \textbf{V}_i )
\end{gather}
Consequently, the MFA enables the model to discern which factors the user prioritizes and which ones are less relevant during their decision-making. For item modeling, the roles of users and items are reversed, treating items as queries, values, and users as keys, to capture item-centric factor importance.

\subsection{Aligning GCN and MFA-Profile Representations}
To integrate the Multi-Factor Attentive embeddings and graph embeddings for recommendation, we adopt a straightforward concatenation approach, where [; ] denotes the concatenation. Despite its simplicity, this method effectively combines the user-item collaborative signals captured by GCN and the semantic preferences represented in the weighted profile embeddings. To ensure seamless integration, the vector dimensions of the two representations are set to be identical. Finally, the matching scores are computed from the inner product of the final user and item embeddings for Top-\(K\) recommendations.
\begin{gather}
    e_u = [e_u^g; e_u^a], \quad e_i = [e_i^g; e_i^a] ; \quad
    \hat{y}_{ui} = e_u^T e_i
\end{gather}
For model training, we optimize the Bayesian Personalized Ranking (BPR) loss \cite{bpr}, a pairwise ranking loss that encourages observed interactions to have higher scores than unobserved ones.
\begin{equation}\label{eq:bpr_loss}
    \mathcal{L}_{\text{BPR}} = 
    - \sum_{(u,i,j) \in O}^M 
    \ln \sigma (\hat{y}_{ui} - \hat{y}_{uj}) 
    + \lambda \|\Theta\|_2^2
\end{equation}Here, \( O = \{ (u,i,j)|(u,i) \in \mathcal{R^+},(u,j) \in \mathcal{R^-} \} \) denotes the pairwise training data. \(\mathcal{R^+}\) and \(\mathcal{R^-}\) indicate the observed and unobserved interactions respectively; \( \sigma(\cdot) \) is the sigmoid function. Additionally, \( \lambda \) controls the \( L_2 \) regularization and \( \Theta\ = \{ \mathbf{E^{(0)}}, \textbf{W}_Q, \textbf{W}_K, \textbf{W}_V\} \) denote all trainable parameters,  where \(\mathbf{E^{(0)}}\) are the embeddings of the 0-th GCN layer.

\section{Experiments}

Here, we conduct experiments to address the following research questions:
\begin{itemize}[leftmargin=1em]
    \item \textbf{RQ1}: How does the proposed framework perform compared to state-of-the-art baseline methods?
    \item \textbf{RQ2}: How do different hyperparameters impact the overall performance of the framework?
    \item \textbf{RQ3}: How do the individual components of the framework contribute to its overall performance?
    \item \textbf{RQ4}: Do the selected factors demonstrate validity in the context of restaurant recommendation?
    \item \textbf{RQ5}: Does LLM-based profiling from user reviews provide discriminative information for capturing user preferences?
\end{itemize}

\subsection{Experimental Setting}


\subsubsection{Datasets.} 
We evaluate on two public datasets providing abundant textual feedback for profile construction: Yelp and Google Restaurants (GR) \cite{google_rst}. Following \cite{rlmrec}, we apply 10-core (Yelp) and 5-core (GR) filtering and partition the data into training, validation, and test sets at a 3:1:1 ratio.

\subsubsection{Evaluation Protocols and Metrics.} 
We measure top-$K$ recommendation performance using Recall@$K$ and NDCG@$K$ ($K \in \{10, 20\}$) under an all-ranking strategy \cite{lgcn,sgl}. Results are averaged over five random seed runs, and statistical significance is assessed through paired t-tests against the best baseline, reporting the corresponding $p$-values.


\subsubsection{Baselines.}
We compare ReFORM with four groups of baselines: (i) GCN Methods: GCCF \cite{gccf} and LightGCN \cite{lgcn}; (ii) Self-supervised GCN Methods: SGL \cite{sgl} and SimGCL \cite{simgcl}; (iii) Contents-based GCN Methods: MMGCN \cite{mmgcn} and GRCN \cite{grcn}; and (iv) LLM-Enhanced GCN: RLMRec \cite{rlmrec}. For fair comparison, we use RLMRec's best-performing contrastive learning variant with an identical backbone model.

\subsubsection{Implementation Details.}
Models are optimized using the Adam optimizer with a $1 \times 10^{-3}$ learning rate, a 4096 batch size, and early stopping on validation. All models use an embedding size of 256, and text-based baselines share our semantic representations. For the Multi-Factor Attention module, we tune the attention keys $n \in \{1, 2, 3, 4, 5\}$.


\subsection{Overall Performance (RQ1)}
Our proposed framework demonstrates consistent and significant improvements over all baseline methods, as shown in Table \ref{table:main_table}. By incorporating our two main methodologies Review-aggregated Profile Generation (RPG) and Multi-Factor Attention (MFA), our model effectively captures detailed user and item preferences at a granular level, resulting in superior recommendation performance.

\begin{table*}[!htbp]
    \caption{Performance comparison between our framework and baselines on two datasets. The best results are highlighted in bold, and the second bests are underscored. The relative improvements compared to the best baselines are indicated as Improv.}
    \centering
     \scalebox{0.95}{
    \begin{tabular}{lcccccccccc}
    \toprule[1pt]
    \multirow{2}{*}{Baseline}   & \multicolumn{4}{c}{\textbf{Yelp}}   & &  & \multicolumn{4}{c}{\textbf{Google Restaurants}}  \\ \cmidrule{2-5} \cmidrule{8-11} 
                                & R@10 & R@20 & N@10 & N@20 & &  & R@10 & R@20 & N@10 & N@20 \\
    \midrule
    & \multicolumn{10}{c}{Graph Convolution Networks (GCN) Methods} \\ \midrule 
    GCCF                        & 0.0390	  & 0.0643	   & 0.0331	 & 0.0414    & &    & 0.0570     & 0.0925   & 0.0351     & 0.0457    \\
    LightGCN                    & 0.0421     & 0.0702     & 0.0357     & 0.0450    & &	 & 0.0574	  & 0.0931	   & 0.0357    & 0.0464    \\ \midrule
    & \multicolumn{10}{c}{Self-supervised GCN Methods}         \\ \midrule
    SGL                         & 0.0485     & 0.0806     & 0.0409     & 0.0516   & & & 0.0583 & 0.0932 & 0.0362 & 0.0468   \\
    SimGCL                      & 0.0389     & 0.0645     & 0.0331     & 0.0416   & & & \underline{0.0607}	  & \underline{0.0973}	   & \underline{0.0377}	 & \underline{0.0487}    \\ \midrule

    & \multicolumn{10}{c}{Contents-based GCN Methods}                \\ \midrule
    MMGCN   & 0.0416 & 0.0714 & 0.0340 & 0.0441 & & & 0.0158 & 0.0268 & 0.0092 & 0.0126 \\
    GRCN    & 0.0509 & 0.0877 & 0.0427 & 0.0551 & & & 0.0486 & 0.0796 & 0.0300 & 0.0394 \\ \midrule
    & \multicolumn{10}{c}{LLM-Enhanced GCN Methods}                \\ \midrule
    RLMRec  & \underline{0.0597} & \underline{0.0973} & \underline{0.0505}     & \underline{0.0630}  &  &  & 0.0598     & 0.0962     & 0.0371  & 0.0480  \\
    
    \textbf{ReFORM}     & \textbf{0.0650}  & \textbf{0.1062}  & \textbf{0.0546}  & \textbf{0.0683}  & &  & \textbf{0.0685}  & \textbf{0.1088}  & \textbf{0.0424}  & \textbf{0.0545}  \\  \midrule
    \(p\)-value.                     & $3.4e^{-4}$     & $3.2e^{-5}$    & $3.0e^{-4}$  &  $4.9e^{-5}$ & &  & $9.0e^{-4}$  & $1.0e^{-4}$   & $1.6e^{-3}$  & $9.0e^{-4}$  \\
    Improv.                     & 8.88\%     & 9.15\%    & 8.12\%   & 8.41\%  & & & 12.85\%  & 11.82\%  & 12.47\%  & 11.91\%  \\
    \bottomrule[1pt]
    \end{tabular}
    }
    \label{table:main_table}
\end{table*}

Integrating factor-specific generated profiles substantially enhances recommendation performance. LLM-enhanced methods, such as RLMRec and ReFORM, which leverage rich profile information, deliver superior recommendation performance compared to Graph Convolution Networks (GCN) methods. ReFORM achieves improvements of 51.3\% and 16.7\% in Recall@20 over its LightGCN backbone on Yelp and Google Restaurants (GR), respectively. Compared with SGL and SimGCL, which learn expressive representations through self-supervised learning, these methods still rely solely on internal collaborative signals. In contrast, ReFORM actively integrates external information to directly capture granular user- and item-specific details. Since SGL, SimGCL, and ReFORM all utilize LightGCN as their backbone, performance gaps directly reflect the effectiveness of their representation enhancement strategies. Specifically, ReFORM achieves a Recall@20 improvement of 51.3\% on Yelp and 16.7\% on GR over LightGCN, far exceeding the gains of the best SSL method (14.8\% and 4.4\%, respectively), clearly underscoring the advantage of integrating rich external information.

Contents-based GCN methods, MMGCN and GRCN, outperform the GCN methods on Yelp but lag behind LLM-enhanced models by relying solely on item-level features without modeling user-side semantics. Unexpectedly, these methods perform poorly on the GR dataset, as we assume it does not provide sufficient interaction and item features due to its small number of reviews. In contrast, ReFORM effectively models the correlation between interactions and textual preferences even on a small dataset by utilizing features at both the user and item levels. Furthermore, ReFORM outperforms RLMRec, a recent LLM-enhanced GCN approach. In particular, ReFORM achieves improvements over RLMRec by 9.2\% and 13\% in Recall@20 on the Yelp and GR datasets, respectively. While RLMRec effectively integrates textual information into graph structure through contrastive learning, it lacks the ability to differentiate the importance of diverse semantic signals. In contrast, ReFORM combines graph-based collaborative signals with Multi-Factor Attentive profiles, effectively adjusting user and item factor preferences over interacted decisions.

\begin{figure*}[!t]
  \centering
  \begin{minipage}{0.48\textwidth}
    \centering
    \includegraphics[width=\textwidth]{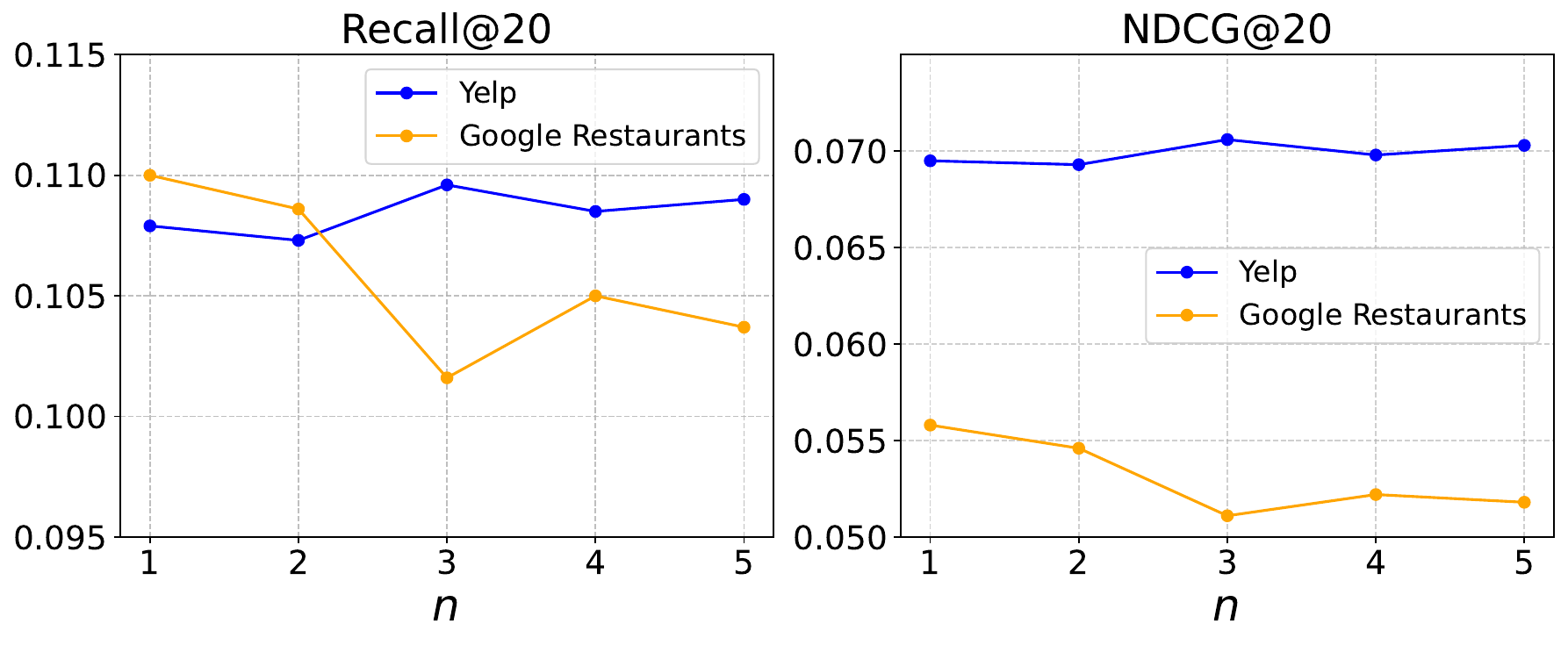}
    \caption{Influence of the number of keys \( n \) of Multi-Factor Attention.}
    \label{figure:hype_n}
  \end{minipage}
  \hfill 
  \begin{minipage}{0.48\textwidth}
    \centering
    \includegraphics[width=\textwidth]{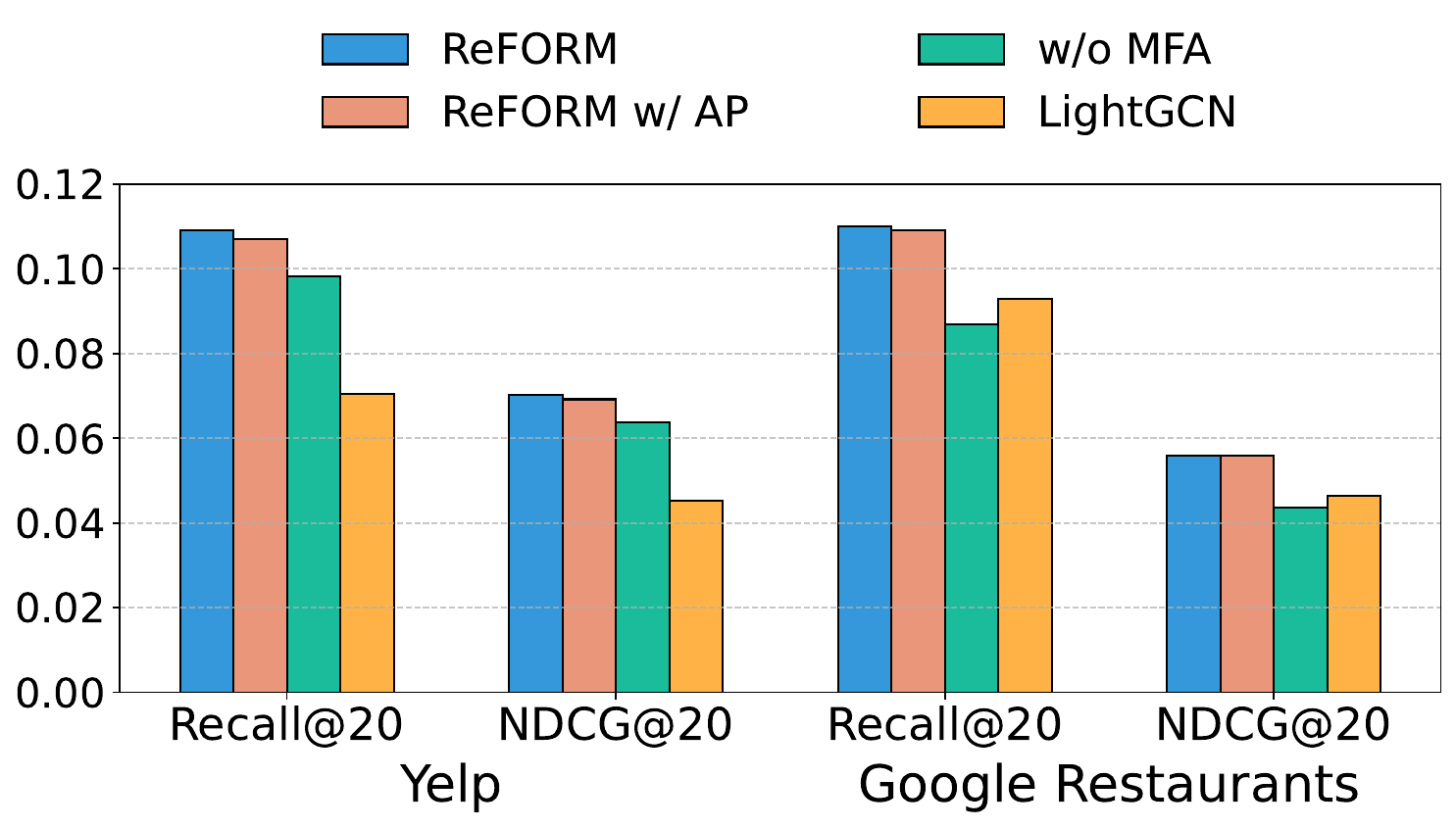}
    \caption{Ablation study on the components of ReFORM framework.}
    \label{figure:abl}
  \end{minipage}
\end{figure*}

\subsection{Hyperparameter Analysis (RQ2)}

In Figure \ref{figure:hype_n}, the number \( n \) of keys \textbf{K} used in the Multi-Factor Attention (MFA) of Eq. \eqref{eq:MFA_formula} controls the tradeoff between bias and overfitting. On the Yelp dataset, we observe that the MFA achieves the best performance at \(n\) = 3. We believe that using fewer \textbf{K} keys (e.g., 1 or 2) introduces high bias, whereas using more keys (e.g., 4 or 5) results in overfitting. On the Google Restaurants (GR) dataset, MFA achieves the best performance at \(n\) = 1 and the results decline in performance as the number of keys increases. We assume that this is because GR has lower average user interactions than the Yelp, which makes the data sparse. Sparse records increase the likelihood of sampling the same keys repeatedly across epochs, which leads to overfitting and ultimately poor performance.

\subsection{Ablation Study (RQ3)}

To comprehensively analyze the contributions of key components in the proposed framework, we conduct an ablation study. The results, presented in Figure \ref{figure:abl}, demonstrate the importance of the framework’s key modules. 


In Multi-Factor Attention (MFA), attention maps from \( n \) keys are aggregated using max pooling to accentuate factor importance. We replaced max pooling with average pooling (\textit{ReFORM w/ AP}) to compare the aggregation methods. As shown in Figure \ref{figure:abl}, \textit{ReFORM w/ AP} demonstrates slightly lower performance compared to \textit{ReFORM}, suggesting that average pooling may dilute the emphasis on dominant preferences. This finding highlights the efficacy of max pooling in preserving strong factor preference during the MFA process.

Multi-Factor Attention (MFA) adjusts influential factors at both the user and item levels to optimize their representations. We replaced this mechanism with a simpler MLP, denoted as \textit{w/o MFA}, to remove the factor preference adjustment step. This substitution led to a significant performance drop, highlighting the essential role of MFA in adjusting factor preferences for improved representation learning. Notably, on the Google Restaurants dataset, removing MFA even degrades performance below that of the base model, \textit{LightGCN}. This suggests that using factor-specific information without proper adjustment may introduce noise, negatively impacting the learning process \cite{mlp_noise}. These findings highlight the critical role of the MFA mechanism in capturing nuanced user and item preferences effectively.


\section{In-depth Analysis of ReFORM}

\begin{table*}[t]
    \centering
    \caption{Ablation study on factor importance for Yelp and Google Restaurants datasets. $\Delta$ (\%) denotes the performance change compared to ReFORM.}
    \begin{tabular}{lcccccccccc}
        \toprule
        \multirow{2}{*}{Method} 
            &
            & \multicolumn{4}{c}{\textbf{Yelp}} 
            &
            & \multicolumn{4}{c}{\textbf{Google Restaurants}} \\
        \cmidrule(lr){3-6} \cmidrule(lr){8-11}
            &
            & R@20 & $\Delta$ (\%) & N@20 & $\Delta$ (\%) 
            &
            & R@20 & $\Delta$ (\%) & N@20 & $\Delta$ (\%) \\
        \midrule
        \textbf{ReFORM}         & & 0.1062 & - & 0.0683 & - & & 0.1088 & - & 0.0545 & - \\
        w/o cuisine type        & & 0.0982 & \textcolor{red}{-7.53} & 0.0634 & \textcolor{red}{-7.17} & & 0.0932 & \textcolor{red}{-14.34} & 0.0468 & \textcolor{red}{-14.13} \\
        w/o flavor              & & 0.1032 & \textcolor{red}{-2.82} & 0.0664 & \textcolor{red}{-2.78} & & 0.1004 & \textcolor{red}{-7.72} & 0.0505 & \textcolor{red}{-7.34} \\
        w/o atmosphere          & & 0.1055 & \textcolor{red}{-0.66} & 0.0679 & \textcolor{red}{-0.59} & & 0.1030 & \textcolor{red}{-5.33} & 0.0519 & \textcolor{red}{-4.77} \\
        w/o price               & & 0.1060 & \textcolor{red}{-0.19} & 0.0681 & \textcolor{red}{-0.29} & & 0.1069 & \textcolor{red}{-1.75} & 0.0538 & \textcolor{red}{-1.28} \\
        w/o companion           & & 0.1061 & \textcolor{red}{-0.09} & 0.0682 & \textcolor{red}{-0.15} & & 0.1078 & \textcolor{red}{-0.92} & 0.0544 & \textcolor{red}{-0.18} \\
        w/o time                & & 0.1044 & \textcolor{red}{-1.69} & 0.0671 & \textcolor{red}{-1.76} & & 0.1081 & \textcolor{red}{-0.64} & 0.0545 & \textcolor{blue}{0.00} \\
        w/o waiting             & & 0.1060 & \textcolor{red}{-0.19} & 0.0682 & \textcolor{red}{-0.15} & & 0.1086 & \textcolor{red}{-0.18} & 0.0549 & \textcolor{blue}{0.73} \\
        \bottomrule
    \end{tabular}
    \label{table:factor_validity}
\end{table*}

\subsection{Validation of Selected Factors (RQ4)}

In this section, we conduct an in-depth analysis of the selected factors in the restaurant recommendation setting. We evaluated factor importance through an ablation experiment, in which each factor-specific profile embedding is masked in turn to quantify its contribution to recommendation performance.

To assess the importance of each selected factor, we conduct an ablation study by individually masking the text-derived profile embedding corresponding to each factor during inference. Specifically, for each factor, we replace the corresponding slice in the user and item profile embeddings with a zero vector, thereby removing the contribution of that factor in the recommendation process. We then evaluate the absolute and relative changes in performance on the test set. The results of this experiment are summarized in Table \ref{table:factor_validity}, which quantifies the contribution of each factor in the recommendation performance. Notably, ablating \textit{cuisine type} and \textit{flavor} causes the most severe performance drops across both datasets (e.g., R@20 drops by 14.34\% and 7.72\% in Google Restaurants, respectively), demonstrating their critical roles in capturing personalized preferences. Interestingly, \textit{atmosphere} shows a marked impact on Google Restaurants (R@20: -5.33\%) but a limited effect on Yelp (-0.66\%), indicating subtle platform-specific differences in user priorities. Conversely, removing \textit{time} and \textit{waiting} yields minimal reductions or even slight improvements. This implies that these peripheral factors are less informative and may occasionally introduce noise, diluting the model's focus. Overall, these results validate the effectiveness of the selected factors, providing actionable insights for both future model design and real-world system deployment.

\subsection{Noise Review Injection in RPG (RQ5)}

\begin{table}[t]
    \centering
    \caption{Impact of noise ratio in RPG profile generation on Google Restaurants dataset. 
    $\Delta$ (\%) denotes the relative performance change compared to the original ReFORM (noise ratio = 0).}
    \begin{tabular}{l|cc|cc}
        \toprule
        \textbf{Noise ratio}  & \textbf{R@20}   & \textbf{$\Delta$ (\%)} & \textbf{N@20}   & \textbf{$\Delta$ (\%)} \\
        \midrule
        0                     & \textbf{0.1088} & --                     & \textbf{0.0545} & -- \\
        0.5                   & 0.0962          & \textcolor{red}{-11.58} & 0.0481          & \textcolor{red}{-11.74} \\
        1.0                   & 0.0915          & \textcolor{red}{-15.9} & 0.0457          & \textcolor{red}{-16.15} \\
        \bottomrule
    \end{tabular}
    \label{table:rpg_noise}
\end{table}

To evaluate the robustness of Review-aggregated Profile Generation (RPG), we systematically introduce irrelevant reviews (noise) into the profile construction for the Google Restaurants dataset. Here, a noise ratio of 0 uses only the target user's authentic reviews, while 1.0 relies entirely on randomly sampled reviews from other users. In Table \ref{table:rpg_noise}, performance degrades monotonically as the noise ratio increases. Even partial contamination (ratio = 0.5) significantly impairs recommendation quality, and complete noise (ratio = 1.0) leads to substantial drops of up to 16.15\% in NDCG@20. This confirms that authentic user reviews provide crucial personalized signals that LLM-powered RPG effectively extracts, underscoring the necessity of high-quality input data for profile generation.

\section{Conclusion}
In this paper, we propose ReFORM, a novel recommendation framework that leverages a large language model to distinguish individual preferences. ReFORM generates factor-specific user/item profiles from reviews only via Review-aggrega\-ted Profile Generation (RPG). By introducing Multi-Factor Attention (MFA), our framework captures the factor-specific preferences of users and items, that provides understanding of which factors drive decision-making. We demonstrate the robustness of our framework by evaluating ReFORM on restaurant datasets of different sizes, where it consistently outperforms state-of-the-art baselines. Moving forward, we plan to further extend ReFORM’s capabilities by exploring other domains. We also aim to apply advanced graph modeling (e.g., contrastive learning) and other recommendation paradigms (e.g., sequential recommendation) to deepen our insights into user behavior and improve the practical usability of the framework.



\bibliographystyle{splncs04}
\bibliography{references}
\end{document}